\documentclass[aip,reprint,amssymb,amsmath,superscriptaddress,showpacs,twocolumn]{revtex4-1}
\usepackage{graphicx}
\usepackage{dcolumn}
\usepackage{bm}
\begin{document}

\title[P. Perna \emph{et al.}]{Conducting interfaces between band insulating oxides: the LaGaO$_3$/SrTiO$_3$ heterostructure}

\author{P. Perna}
\altaffiliation{p.a.: Instituto Madrile\~{n}o de Estudios Avanzados en Nanociencia IMDEA-Nanociencia, Campus Universidad Aut\'{o}noma de Madrid, 28049 Madrid, Spain}
\affiliation{CNR-SPIN and Dipartimento di Scienze Fisiche Univ. di Napoli "Federico II", Compl. Univ. di Monte S.Angelo, Via Cinthia, 80126, Napoli, Italy}
\author{D. Maccariello}
\author{M. Radovic}
\altaffiliation{p.a.: Swiss Light Source, Paul Scherrer Institut, WSLA 006, CH-5232 Villigen PSI, Switzerland}
\author{U. Scotti di Uccio}
\affiliation{CNR-SPIN and Dipartimento di Scienze Fisiche Univ. di Napoli "Federico II", Compl. Univ. di Monte S.Angelo, Via Cinthia, 80126, Napoli, Italy}
\author{I. Pallecchi}
\author{M. Codda}
\author{D. Marr\'{e}}
\affiliation{CNR-SPIN and Dipartimento di Fisica, Universita' di Genova, via Dodecaneso 33, 16146, Genova, Italy}
\author{C. Cantoni}
\author{J. Gazquez}
\author{M. Varela}
\author{S.J. Pennycook}
\affiliation{Materials Science and Technology Division, Oak Ridge National Laboratory, 1 Bethel Valley Road, Oak Ridge, TN 37831-6116, USA}
\author{F. Miletto Granozio}
\altaffiliation{Corresponding author: e-mail miletto@na.infn.it}
\affiliation{CNR-SPIN and Dipartimento di Scienze Fisiche Univ. di Napoli "Federico II", Compl. Univ. di Monte S.Angelo, Via Cinthia, 80126, Napoli, Italy}
\date{\today ~---~ Appl.~Phys.~Lett.~\textbf{97}, 1 (2010) \textit{in press}}

\begin{abstract}
 We show that the growth of the heterostructure LaGaO$_{3}$/SrTiO$_{3}$ yields the formation of a highly conductive interface. Our samples were carefully analyzed by high resolution electron microscopy, in order to assess their crystal perfection and to evaluate the abruptness of the interface. Their carrier density and sheet resistance are compared to the case of LaAlO$_{3}$/SrTiO$_{3}$ and a superconducting transition is found. The results open the route to widening the field of polar-non polar interfaces, pose some phenomenological constrains to their underlying physics and highlight the chance of tailoring their properties for future applications by adopting suitable polar materials.
\end{abstract}


\maketitle

The quasi 2-dimensional electron gas (q2-DEG) recently discovered at the LaAlO$_3$ (LAO)/SrTiO$_3$ (STO) interface\cite{othomo} is presently envisaged as an ideal system for the realization of nanoscale oxide devices.\cite{cen} The \textit{electronic reconstruction} model attributes the origin of the q2-DEG to an electronic relaxation mechanism occurring at the interface between the (nominally) non-polar (001) STO substrate and the polar (001) LAO film. The wide band-gap of LAO is considered as crucial in this approach, because it determines the capability of the polar film to transfer charges over the band gap of STO. Ideally, half an electron per areal unit cell ($\approx 3.3 \times 10^{14} \text{cm}^{-2}$) is expected to be transferred at the TiO$_{2}$-LaO interface, partially filling the 3$d$ Ti levels of the STO conduction band (CB). Alternatively, a possible active role of oxygen vacancies in STO near the interface was envisaged.\cite{siemons} Actually, the transport properties of the heterostructure are affected both by oxygen pressure during growth \cite{brinkman, basletic} and by the application of an oxygen post-anneal.\cite{basletic} Finally, it was argued that a substantial La substitution for Sr during sample growth might drive the insulating surface of STO into a conductor.\cite{Goniakowski, willmott, willmott2}
Obviously, also LAO poses material issues.\cite{pentcheva} In this context, we started the search of novel heterostructures based on a different overlayer. On this basis, we identified as a first test material LaGaO$_3$ (LGO), a polar, wide band gap, pseudocubic perovskite.

\begin{figure}[t]
\includegraphics[scale=0.45]{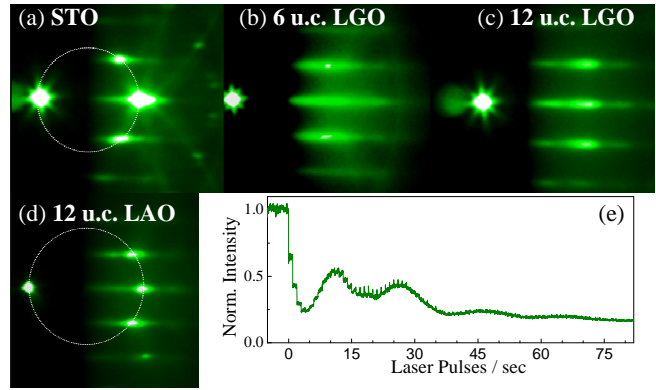}\\

  \caption{(color online). The RHEED pattern of a (a) STO single crystal, (b) 6 u.c.~thick and (c) 12 u.c.~thick LGO film surface compared with a (d) 12 u.c.~thick LAO film. An example of RHEED oscillations during the growth of LAO is shown in (e).}\label{Fig_1}
\end{figure}

Films of LAO and LGO were deposited on nominally TiO$_2$ terminated STO substrates, chemically treated in de-ionised water and buffered-HF.\cite{radovic_SrO,koster} The growth was performed by Reflection High Energy Electron Diffraction (RHEED) assisted Pulsed Laser Deposition (KrF excimer laser, $248$nm) with a typical fluence of $\approx 1.5 - 2.5$ J cm$^{-2}$ at the target, a substrate temperature of $800^{\circ}$C and different oxygen pressures within the $10^{-2} - 10^{-4}$ mbar range.\cite{NB} LAO films presented regular RHEED oscillations typical of layer-by-layer growth and a final pattern reminiscent of a single crystal surface, whereas LGO films showed damped and less regular oscillations, and a streaky 2D pattern at the end of the growth (Fig.~\ref{Fig_1}).

\begin{figure}
\includegraphics[scale=0.45]{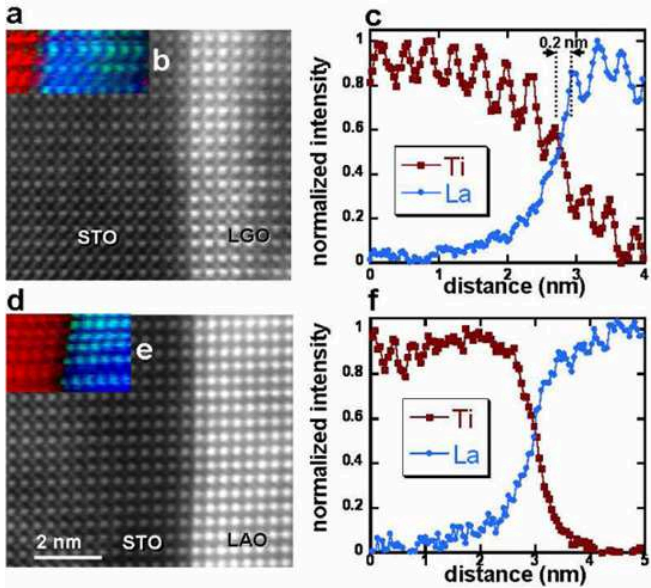}\\

  \caption{(color online). ADF-STEM images of (a) the LGO/STO and (d) the LAO/STO interface, grown in an oxygen partial pressure of $10^{-4}$ mbar and $10^{-3}$ mbar respectively, both viewed down the STO {100} zone axis. The La ions are the brightest followed by the Sr ions. The Ti ions appear smaller and dimmer than the Sr ions. The contrast is insufficient to resolve O, Al, and Ga ions. Plots (c) and (f) show the result of EELS line scans. Plotted are the integrated intensities of the $Ti-L_{2,3}$ (squares) and the $La-M_{4,5}$ (circles) transition edges. The maxima correspond to the approximate position of Ti and La columns respectively. Insets (b) and (e) are Ti and La maps obtained on LGO/STO and LAO/STO respectively through acquisition of EEL spectrum images. Square red symbols indicate Ti columns and light blue circles indicate La columns. }\label{Fig_2}
\end{figure}

The atomic and electronic structures of LAO/STO and LGO/STO interfaces were investigated by high-resolution scanning transmission electron microscopy (STEM) and electron energy loss spectroscopy (EELS) measurements performed in an aberration-corrected VG Microscope $HB501UX$ operated at $100$ kV and equipped with an Enfina EEL spectrometer and a Nion aberration corrector. Cross section imaging of the samples at several locations revealed that the LAO and LGO films were essentially defect-free and had grown with a \textit{cube-on-cube} epitaxial relationship of their pseudocubic cell on the STO cell. Selective area diffraction patterns acquired in a Philips CM200 (FEG 200 kV) TEM microscope did not show the reflections from the orthorhombic LGO phase that are related to the antiferrodistortive order of the octahedra. The interface between film and substrate appeared atomically flat and coherent, and no dislocations or strain fields indicating film relaxation were present. An analysis of the relative lattice parameters of film and substrate based on images and diffraction patterns confirmed that our LGO and LAO films were single-domain and tetragonally strained with an estimated in-plane lattice constant equal to that of the STO substrate and an estimated out-of-plane lattice constant equal to ($3.86 \pm 0.04$) $\AA$ and ($3.71 \pm 0.04$) $\AA$ for LGO and LAO respectively. Cation intermixing at the interface was investigated by atomically resolved EELS spectrum images and line scans.\cite{varela} Fig.~\ref{Fig_2}(a) and Fig.~\ref{Fig_2}(d) show high resolution Z-contrast STEM images of the LGO/STO and LAO/STO interfaces with the La columns showing the brightest contrast due to the large La atomic number.\cite{pennycook, varela2} The insets labelled (b) and (e) are atomically resolved EEL spectrum images of an interface region. The maps were obtained by merging the $Ti-L$ (red) and $La-M$ (light blue) spectroscopic signals and suggest, in spite of the presence of a minor sample drift, an atomically sharp interface. The sequence of atomic planes at the interface was further investigated by acquiring additional EELS line profiles with larger beam dwell time to better resolve fine structure details of the $Ti-L$ and $O-K$ edges. Fig.~\ref{Fig_2}(c) and Fig.~\ref{Fig_2}(f) show the result of such scans through plots of the integrated Ti-L$_{2,3}$ and La-M$_{4,5}$ intensities as a function of distance. Ripples corresponding to the discrete atomic columns distribution can be clearly seen in both the Ti and La profiles, proving the high spatial resolution of the EEL spectra. In the LGO/STO sample the La signal decays from about $85\%$ of the bulk signal on the first LaO monolayer to about $30\%$ in the last SrO monolayer of the substrate. In the LAO/STO sample the La signal decays from $75\%$ of the bulk signal on the first LaO monolayer to $22\%$ in the last SrO monolayer. Given the experimental broadening of the EELS signal, mainly arising from inelastic electron scattering, both profiles are compatible with atomically abrupt interfaces and sharper than the data reported in Refs. \cite{willmott} and \cite{Goniakowski}. The existence of a Ti signal past the interface on the La side of the graph for the LGO/STO sample could be due to a minor Ti interdiffusion into the LGO or, more likely, to Ti atoms sputtered on the specimen surface as a result of sample preparation. The latter conclusion is supported by the observation that the Ga profile is sharper than the Ti profile and does not suggest interdiffusion of Ga into the STO (not shown here). The oscillations in the Ti signal on the LGO side might reflect, as is often observed, the periodicity of the underlying LGO crystal rather than the periodicity of surface Ti atoms. We notice that the spacing between the first La peak (indicating the position of the first LaO plane) and the preceding Ti peak (indicating the position of the last substrate's TiO$_2$ plane) is about half a STO unit cell (Fig.~\ref{Fig_2}(c)). This observation is consistent with an expected interface plane sequence of the type GaO$_{2}$/LaO/TiO$_{2}$ also known as n-type interface.  The LAO/STO samples also showed an n-type interface with the plane sequence AlO$_{2}$/LaO/TiO$_{2}$, as illustrated by the EELS map in the inset of Fig.~\ref{Fig_2}(d).

Transport properties measurements (magnetoresistance, Hall effect and Hall mobility) were carried out on a batch of samples in a PPMS Quantum design system from $2.5$K to room temperature and in magnetic fields up to $9$T.
All samples discussed below were grown on carefully etched and treated STO. A proper surface preparation proved to be of paramount importance, since we could verify that an accidental inaccurate treatment on a batch of substrates, revealed by RHEED and attributed to partial double termination, resulted in the growth of insulating interfaces with room temperature resistance of about $4-8$ M$\Omega $. The measurements were performed in a standard Hall bar geometry resorting to ultrasonic bonding of the contacts.
Room temperature conductivity values were in the range of 10 k$\Omega$s for LAO/STO and LGO/STO samples with a thickness in the range 4-12 u.c. The threshold thickness for the formation of the q-2DEG in our LAO/STO samples had been previously analysed in detail,\cite{savoia} by complementing transport data with optical second harmonic generation measurements. Interestingly, our present data show that the onset of conductivity takes place at ~4 u.c. also for the LGO/STO interface. Without drawing any conclusion, we remark that LGO and LAO also share similar values of the dielectric constant ($ \epsilon \approx$ 25)\cite{dube}, i.e. the principal parameter determining the polar layer response within the “electronic reconstruction” model. \cite{thiel,pentcheva}

\begin{figure}
\includegraphics[scale=0.8]{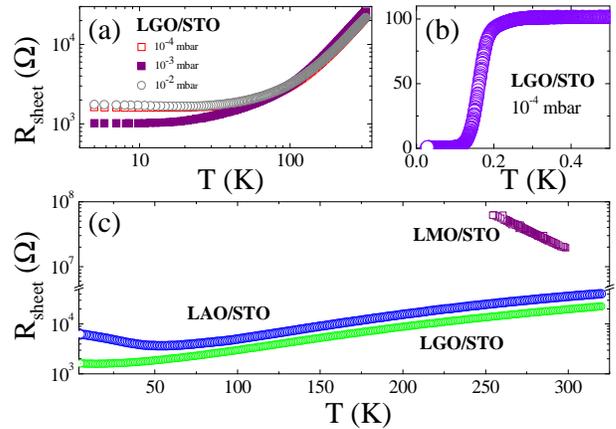}\\

  \caption{(color online). The sheet resistance of three LGO/STO interfaces grown at different oxygen pressure is shown in (a). The transition to the superconducting state of a LGO/STO sample grown at $10^{-4}$ mbar is shown in (b). The sheet resistance versus temperature of 12 u.c.~LAO, LGO and LMO based interfaces grown at the pressure of $10^{-4}$ mbar is reported in (c). }\label{Fig_3}
\end{figure}

The R(T) curves of LGO/STO and LAO/STO are qualitatively similar, except for the low temperature behaviour: LAO/STO shows a resistivity upturn, as reported for samples grown in a similar pressure range,\cite{brinkman} while LGO/STO is characterized by a lower residual resistivity (Fig.~\ref{Fig_3}(a)) that has no clear relation with the oxygen pressure during growth up to $10^{-2}$ mbar. The Hall effect measurements indicate both for LGO/STO and for LAO/STO a sheet carrier density in the range $1 \times 10^{14} - 3 \times 10^{14}$ cm$^{-2}$ at room temperature, with a weak decrease at low temperature. The Hall mobility exhibits the typical T$^{-2}$ dependence at high temperature and saturates at low temperature to values in the range $10^{-3} - 10^{-2}$ m$^{2}$s$^{-1}$V$^{-1}$. Quite interestingly, the LGO-based interfaces show the highest values.
Mobility values where also extracted from magnetoresistance data. $\mu_{\text{MR}}$ also exhibits a power low behaviour above 100 K and a saturation at low temperature.
In some samples, $\mu_{\text{MR}}$ and $\mu_{\text{Hall}}$ values differ from each other by a factor 10 or more at low temperature. This discrepancy is commonly found in these systems\cite{bell} and ascribed to additional scattering mechanisms in different temperature ranges.

The LGO/STO interfaces are superconducting at low temperature (Fig.~\ref{Fig_3}(b)) with a critical temperature $T_C$ of about $150$ mK, as previously reported for LAO/STO.\cite{reyren}

As a second test material, we grew LaMnO$_3$ (LMO) over STO at the same deposition condition of LGO and LAO. A sample with 12 u.c.~thick LMO showed an insulating behaviour, with a resistance which became immeasurably high below  250 K, which we attribute to the intrinsic resistivity of LMO (Fig.~\ref{Fig_3}(c)). This result demonstrated, at least, that having a La-based, polar perovskite grown on Ti terminated STO is not a sufficient condition for the q2-DEG formation.
It is worth mentioning that LMO/STO showed an insulating behavior for growth pressures as low as P(O$_{2}$) $=10^{-4}$ mbar, while conductive LAO/STO and LGO/STO samples were grown up to P(O$_{2}$) $=10^{-2}$ mbar. Obviously, this is not a prove that the mechanism of conductivity in LGO/STO is independent of the oxygen content; however, it is hard to hypothesize that LAO and LGO do attract oxygen vacancies close to the interface (e.g., at the terminating TiO$_2$ layer), while LMO does not. We believe that either the magnetic moment of Mn$^{3+}$ ions or the difference between a wide bandgap insulator and a Mott insulator may play a role. We remind however that the observation of a q2-DEG at the interface between STO and the Mott insulator LaVO$_3$ has been reported,\cite{Hwang1}, although the growth at very low oxygen pressure (P(O$_{2}$) $=10^{-6}$ mbar) and specific material issues \cite{Hwang2} set some differences between the two cases.

In conclusion, our work clearly demonstrates that other polar oxides, beyond LAO, can be used to generate a q2-DEG at the interface with STO. The case of LGO was investigated in detail, proving that excellent epitaxy can be achieved under suitable conditions, with EELS profiles compatible with atomically abrupt interfaces. The high crystal perfection of LGO and of its interface with STO is a result in itself, since previous reports indicated poor epitaxy.\cite{shibuya}. Finally, the counterexample of the heterostructure based on the LMO Mott insulator stresses out the complexity of the problem, showing that in nominally very similar conditions (in regard to oxygen vacancy formation, nominal polarity and A-site cation) no interface conductivity is obtained. Our data suggest that identifying the systems that do and do not give rise to the formation of a Q-2DEG might well the most straightforward route to definitively single out the origin of interface conductivity in polar-non polar heterostructures.

The authors gratefully acknowledge S. Gariglio, A. Caviglia, C. Cancellieri, N. Reyren, J.-M. Triscone for measuring the superconducting properties of our samples and for the fruitful discussions. I.P., M.C.~and D.M.~acknowledge financial support from EU under the FP6-STREP project NANOXIDE contract 033191. C.C., M.V.~and S.J.P.~acknowledge financial support from the US Department of Energy, Office of Electricity Delivery and Energy Reliability (C.C.) and Division of Materials Sciences and Engineering (M.V. and S.J.P.).


\begin{thebibliography}{99}
\bibitem{NB}
The samples were cooled to room temperature in the same oxygen pressure employed for the deposition.
\end{thebibliography}
\end{document}